


\font\bigbold=cmbx12
\font\ninerm=cmr9
\font\eightrm=cmr8
\font\sixrm=cmr6
\font\fiverm=cmr5
\font\ninebf=cmbx9
\font\eightbf=cmbx8
\font\sixbf=cmbx6
\font\fivebf=cmbx5
\font\ninei=cmmi9  \skewchar\ninei='177
\font\eighti=cmmi8  \skewchar\eighti='177
\font\sixi=cmmi6    \skewchar\sixi='177
\font\fivei=cmmi5
\font\ninesy=cmsy9 \skewchar\ninesy='60
\font\eightsy=cmsy8 \skewchar\eightsy='60
\font\sixsy=cmsy6   \skewchar\sixsy='60
\font\fivesy=cmsy5
\font\nineit=cmti9
\font\eightit=cmti8
\font\ninesl=cmsl9
\font\eightsl=cmsl8
\font\ninett=cmtt9
\font\eighttt=cmtt8
\font\tenfrak=eufm10
\font\ninefrak=eufm9
\font\eightfrak=eufm8
\font\sevenfrak=eufm7
\font\fivefrak=eufm5
\font\tenbb=msbm10
\font\ninebb=msbm9
\font\eightbb=msbm8
\font\sevenbb=msbm7
\font\fivebb=msbm5
\font\tensmc=cmcsc10


\newfam\bbfam
\textfont\bbfam=\tenbb
\scriptfont\bbfam=\sevenbb
\scriptscriptfont\bbfam=\fivebb
\def\Bbb{\fam\bbfam}

\newfam\frakfam
\textfont\frakfam=\tenfrak
\scriptfont\frakfam=\sevenfrak
\scriptscriptfont\frakfam=\fivefrak
\def\frak{\fam\frakfam}

\def\smc{\tensmc}


\def\eightpoint{%
\textfont0=\eightrm   \scriptfont0=\sixrm
\scriptscriptfont0=\fiverm  \def\rm{\fam0\eightrm}%
\textfont1=\eighti   \scriptfont1=\sixi
\scriptscriptfont1=\fivei  \def\oldstyle{\fam1\eighti}%
\textfont2=\eightsy   \scriptfont2=\sixsy
\scriptscriptfont2=\fivesy
\textfont\itfam=\eightit  \def\it{\fam\itfam\eightit}%
\textfont\slfam=\eightsl  \def\sl{\fam\slfam\eightsl}%
\textfont\ttfam=\eighttt  \def\tt{\fam\ttfam\eighttt}%
\textfont\frakfam=\eightfrak \def\frak{\fam\frakfam\eightfrak}%
\textfont\bbfam=\eightbb  \def\Bbb{\fam\bbfam\eightbb}%
\textfont\bffam=\eightbf   \scriptfont\bffam=\sixbf
\scriptscriptfont\bffam=\fivebf  \def\bf{\fam\bffam\eightbf}%
\abovedisplayskip=9pt plus 2pt minus 6pt
\belowdisplayskip=\abovedisplayskip
\abovedisplayshortskip=0pt plus 2pt
\belowdisplayshortskip=5pt plus2pt minus 3pt
\smallskipamount=2pt plus 1pt minus 1pt
\medskipamount=4pt plus 2pt minus 2pt
\bigskipamount=9pt plus4pt minus 4pt
\setbox\strutbox=\hbox{\vrule height 7pt depth 2pt width 0pt}%
\normalbaselineskip=9pt \normalbaselines
\rm}


\def\ninepoint{%
\textfont0=\ninerm   \scriptfont0=\sixrm
\scriptscriptfont0=\fiverm  \def\rm{\fam0\ninerm}%
\textfont1=\ninei   \scriptfont1=\sixi
\scriptscriptfont1=\fivei  \def\oldstyle{\fam1\ninei}%
\textfont2=\ninesy   \scriptfont2=\sixsy
\scriptscriptfont2=\fivesy
\textfont\itfam=\nineit  \def\it{\fam\itfam\nineit}%
\textfont\slfam=\ninesl  \def\sl{\fam\slfam\ninesl}%
\textfont\ttfam=\ninett  \def\tt{\fam\ttfam\ninett}%
\textfont\frakfam=\ninefrak \def\frak{\fam\frakfam\ninefrak}%
\textfont\bbfam=\ninebb  \def\Bbb{\fam\bbfam\ninebb}%
\textfont\bffam=\ninebf   \scriptfont\bffam=\sixbf
\scriptscriptfont\bffam=\fivebf  \def\bf{\fam\bffam\ninebf}%
\abovedisplayskip=10pt plus 2pt minus 6pt
\belowdisplayskip=\abovedisplayskip
\abovedisplayshortskip=0pt plus 2pt
\belowdisplayshortskip=5pt plus2pt minus 3pt
\smallskipamount=2pt plus 1pt minus 1pt
\medskipamount=4pt plus 2pt minus 2pt
\bigskipamount=10pt plus4pt minus 4pt
\setbox\strutbox=\hbox{\vrule height 7pt depth 2pt width 0pt}%
\normalbaselineskip=10pt \normalbaselines
\rm}


\def\pagewidth#1{\hsize= #1}
\def\pageheight#1{\vsize= #1}
\def\hcorrection#1{\advance\hoffset by #1}
\def\vcorrection#1{\advance\voffset by #1}

\newif\iftitlepage   \titlepagetrue               
\newtoks\titlepagefoot     \titlepagefoot={\hfil} 
\newtoks\otherpagesfoot    \otherpagesfoot={\hfil\tenrm\folio\hfil}
\footline={\iftitlepage\the\titlepagefoot\global\titlepagefalse
           \else\the\otherpagesfoot\fi}

\font\extra=cmss10 scaled \magstep0
\setbox1 = \hbox{{{\extra R}}}
\setbox2 = \hbox{{{\extra I}}}
\setbox3 = \hbox{{{\extra C}}}
\setbox4 = \hbox{{{\extra Z}}}
\setbox5 = \hbox{{{\extra N}}}

\def\RRR{{{\extra R}}\hskip-\wd1\hskip2.0
   true pt{{\extra I}}\hskip-\wd2
\hskip-2.0 true pt\hskip\wd1}
\def\Real{\hbox{{\extra\RRR}}}    


\def\ZZZ{{{\extra Z}}\hskip-\wd4\hskip 2.5 true pt{{\extra Z}}}
\def\Zed{\hbox{{\extra\ZZZ}}}       



\def\r{{\frak r}}

\def\Z{{\Zed}}
\def\R{{\Real}}

\def\frac#1#2{{#1\over#2}}

\def\({\left(}
\def\){\right)}
\def\<{\langle}
\def\>{\rangle}

\def\pmb#1{\setbox0=\hbox{$#1$}%
   \kern-.025em\copy0\kern-\wd0
   \kern.05em\copy0\kern-\wd0
   \kern-.025em\raise.0433em\box0 }


\def\abstract#1{{\parindent=30pt\narrower\noindent\ninepoint\openup
2pt #1\par}}


\newcount\notenumber\notenumber=1
\def\note#1
{\unskip\footnote{$^{\the\notenumber}$}
{\eightpoint\openup 1pt #1}
\global\advance\notenumber by 1}


\global\newcount\secno \global\secno=0
\global\newcount\meqno \global\meqno=1
\global\newcount\appno \global\appno=0
\newwrite\eqmac
\def\romappno{\ifcase\appno\or A\or B\or C\or D\or E\or F\or G\or H
\or I\or J\or K\or L\or M\or N\or O\or P\or Q\or R\or S\or T\or U\or
V\or W\or X\or Y\or Z\fi}
\def\eqn#1{
        \ifnum\secno>0
            \eqno(\the\secno.\the\meqno)\xdef#1{\the\secno.\the\meqno}
          \else\ifnum\appno>0
            \eqno({\rm\romappno}.\the\meqno)\xdef#1{{\rm\romappno}.\the=
\meqno}
          \else
            \eqno(\the\meqno)\xdef#1{\the\meqno}
          \fi
        \fi
\global\advance\meqno by1 }


\global\newcount\refno
\global\refno=1 \newwrite\reffile
\newwrite\refmac
\newlinechar=`\^^J
\def\ref#1#2{\the\refno\nref#1{#2}}
\def\nref#1#2{\xdef#1{\the\refno}
\ifnum\refno=1\immediate\openout\reffile=refs.tmp\fi
\immediate\write\reffile{
     \noexpand\item{[\noexpand#1]\ }#2\noexpand\nobreak.}
     \immediate\write\refmac{\def\noexpand#1{\the\refno}}
   \global\advance\refno by1}
\def\semi{;\hfil\noexpand\break ^^J}
\def\nl{\hfil\noexpand\break ^^J}
\def\refn#1#2{\nref#1{#2}}

\def\ann#1#2#3{{\it Ann. Phys.} {\bf {#1}} (19{#2}) #3}
\def\cmp#1#2#3{{\it Commun. Math. Phys.} {\bf {#1}} (19{#2}) #3}

\def\ijmp#1#2#3{{\it Int.  J.  Mod.  Phys.} {\bf A{#1}} (19{#2}) #3}

\def\np#1#2#3{{\it Nucl.  Phys.} {\bf B{#1}} (19{#2}) #3}
\def\pl#1#2#3{{\it Phys.  Lett.} {\bf {#1}B} (19{#2}) #3}

\def\prD#1#2#3{{\it Phys.  Rev.} {\bf D{#1}} (19{#2}) #3}
\def\prl#1#2#3{{\it Phys.  Rev.  Lett.} {\bf #1} (19{#2}) #3}
\def\ptp#1#2#3{{\sl Prog. Theor. Phys.} {\bf {#1}} (19{#2}) #3}
\def\rmp#1#2#3{{\it Rev.  Mod.  Phys.} {\bf {#1}} (19{#2}) #3}

\def\zpc#1#2#3{{\it Z.  Phys.} {\bf C{#1}} (19{#2}) #3}


{

\refn\Wilczek
{F. Wilczek and A. Zee,
\prl{51}{83}{25};
Y.S. Wu and A. Zee,
\pl{147}{84}{325}}

\refn\Bowick
{M.J. Bowick, D. Karabali and L.C.R. Wijewardhana,
\np{371}{86}{417};
D. Karabali, \ijmp{6}{91}{1369}}

\refn\Semenoff
{G.W. Semenoff and P. Sodano,
\np{328}{89}{753}}

\refn\Forte
{S. Forte,
\rmp{64}{92}{193}}

\refn\tsutsuiA
{H. Kobayashi, I. Tsutsui and S. Tanimura,
\np{514}{98}{667}}

\refn\tsutsuiB
{M. Kimura, H. Kobayashi and I. Tsutsui,
\np{527}{98}{624}}

\refn\BalachandranA
{A.P. Balachandran, G. Marmo, B.S. Skagerstam and A. Stern,
\lq\lq Classical Topology and Quantum States\rq\rq, 
World Scientific, Singapore, 1991}

\refn\BalachandranB
{A.P. Balachandran, A. Stern and G. Trahern, \prD{19}{79}{2416}}

\refn\Jackiw
{R. Jackiw, in \lq\lq Current Algebras and Anomalies\rq\rq,
World Scientific, Singapore, 1985}

\refn\henneaux
{M. Henneaux and C. Teitelboim, \lq\lq Quantization 
of Gauge Systems\rq\rq,
Princeton University Press, New Jersey, 1992}


\refn\Bott
{R. Bott and L.W. Tu,
\lq\lq Differential Forms in Algebraic Topology\rq\rq,
Springer-Verlag, New York, 1982}

\refn\OS
{H.\ Otsu, H.\ Sato,
\ptp{91}{94}{1199}; \zpc{64}{94}{177}}

\refn\Deser
{S. Deser, R. Jackiw and S. Templeton,
\prl{48}{82}{975};
\ann{140}{82}{372}}

\refn\DHokerB
{E. D'Hoker,
\pl{357}{95}{539}}

\refn\Witten
{E. Witten, 
\cmp{92}{84}{455}}

\refn\dhoker
{E. D'Hoker and S. Weinberg,
\prD{50}{94}{605};
E. D'Hoker, \np{451}{95}{725}}

}

\def\ve{\vfill\eject}

\def\vecx{{\bf x}}
\def\vecy{{\bf y}}
\def\vecn{{\bf n}}
\def\Tr{{\rm Tr}}




\pageheight{23cm}
\pagewidth{14.8cm}
\hcorrection{0mm}
\magnification= \magstep1
\def\bsk{%
\baselineskip= 16.8pt plus 1pt minus 1pt}
\parskip=5pt plus 1pt minus 1pt
\tolerance 6000


\null

{
\leftskip=100mm
\hfill\break
KEK Preprint 99-19
\hfill\break
May 1999
\hfill\break
\par}

\smallskip
\vfill
{\baselineskip=18pt

\centerline{\bigbold
On Topological Terms}
\centerline{\bigbold in the O(3) Nonlinear Sigma Model}

\vskip 30pt

\centerline{
\smc
Toyohiro Tsurumaru\note
{E-mail:\quad tsuru@tanashi.kek.jp}
\quad {\rm and} \quad
Izumi Tsutsui\note
{E-mail:\quad izumi.tsutsui@kek.jp}
}

\vskip 5pt

{
\baselineskip=13pt
\centerline{\it
Institute of Particle and Nuclear Studies}
\centerline{\it
High Energy Accelerator Research Organization (KEK),
Tanashi Branch}
\centerline{\it Tokyo 188-8501, Japan}
}

\vskip 70pt

\abstract{%
{\bf Abstract.}\quad
Topological terms in the $O(3)$ nonlinear sigma
model in $(1+1)$ and $(2+1)$ dimensions are re-examined 
based on the description of the $SU(2)$-valued field $g$.  
We first show 
that the topological soliton term in $(1+1)$ dimensions
arises from the unitary
representations of the group characterizing 
the global structure of the symmetry inherent in
the description, in a manner 
analogous to the appearance of 
the $\theta$-term in Yang-Mills theory 
in $(3+1)$ dimensions.  
We then present a detailed argument as to why 
the conventional 
Hopf term, which is the topological counterpart 
in $(2+1)$ dimensions and has been widely used to
realize fractional spin and statistics,
is ill-defined unless
the soliton charge vanishes.  We show how this
restriction can be lifted by means of a procedure
proposed recently, and provide its
physical interpretation as well.
}

\bigskip
{\ninepoint
PACS codes: 11.10.Kk; 11.27.+d; 03.65.-w \hfill\break
}


\pageheight{23cm}
\pagewidth{15.7cm}
\hcorrection{-1mm}
\magnification= \magstep1
\def\bsk{%
\baselineskip= 14.6pt plus 1pt minus 1pt}
\parskip=5pt plus 1pt minus 1pt
\tolerance 8000
\bsk

\ve

\secno=1 \meqno=1


\noindent{\bf 1. Introduction}
\medskip

The $O(3)$ nonlinear sigma model (NSM) is a model 
ubiquitous in physics, being used in fields ranging from 
condensed matter physics to high energy physics.  It
describes physical systems that 
undergo a spontaneous breakdown of a global 
symmetry $O(3)$ by means of a vector
field $\vecn(\vecx, t)$ with unit length 
$|\vecn|^2 = \sum_{a=1}^3 (n^a)^2 = 1$.  
The dynamics 
of the model is governed by the Lagrangian,
$$
{\cal L}_0(\vecn)=\frac1{2\lambda^2}\ 
(\partial_\mu \vecn)^2,
\eqn\onlsm
$$
where $\lambda$ is a coupling constant and $\mu$
runs from $0$ to the spacetime dimension $d+1$.
It is customary to assume that the field approaches
to a constant vector $\vecn_0$ at spatial infinity,
$$
\vecn(\vecx,t)\to\vecn_0 
\qquad {\rm as}\quad |\vecx|\to\infty.
\eqn\BCone
$$
Due to 
this boundary condition (\BCone), at a fixed time
$t$ the field $\vecn(\vecx, t)$ can be regarded as 
a map from the (effective) space $S^d$
to the target $S^2$ with the fixed value $\vecn_0$ assigned to 
the image of spatial infinity.  In other words, 
the configuration space ${\cal C}_d$
of the model is given by the set of based maps from
$S^d$ to $S^2$, {\it i.e.}, 
${\cal C}_d = {\rm Map}_0(S^d, S^2)$.
  
It has been widely known that the $O(3)$ NSM admits
topological terms which can be added to (\onlsm). 
The best known topological term is the soliton term
in $(1+1)$-dimensions, 
$$
{\cal L}_{\rm soliton} =    
\frac{\hbar\theta}{8\pi}\ \epsilon^{\mu\nu}\epsilon_{abc}
n^a\partial_\mu n^b\partial_\nu n^c\ ,
\eqn\slterm
$$
which is formed out of the volume element of the target space
$S^2$.
The presence of $\hbar$, along with 
the angle parameter $\theta$, signals the fact 
that the term is of quantum origin. 
On the other hand,
in $(2+1)$ dimensions one has the Hopf term,
which has been used to bestow 
fractional spin and statistics upon instanton (skyrmion)
excitations [\Wilczek, \Bowick, \Semenoff, \Forte].
In terms of the field strength 
$
F_{\mu\nu} := 
  - \epsilon^{abc}n^a\partial_\mu n^b\partial_\nu n^c
$
and the connection $A_\lambda$ given as a solution to
$F_{\mu\nu} = \partial_\mu A_\nu - \partial_\nu A_\mu$, 
the conventional Hopf term used in the literature
takes the form of the Chern-Simons term,
$$
{\cal L}_{\rm Hopf} =    
 -\frac{\hbar\theta}{32\pi^2}\ \epsilon^{\mu\nu\lambda}
F_{\mu\nu} A_\lambda,
\eqn\hopfterm
$$
where again $\theta$ is an angle parameter.  

The two topological terms mentioned above share
the origin in that they arise from the same 
topological structure characterized by the 
fundamental group of the respective 
configuration space, 
$$
\pi_1({\cal C}_1)=\pi_2(S^2)=\Z\ , 
\qquad
\pi_1({\cal C}_2)=\pi_3(S^2)=\Z\ ,
\eqn\firsthomotopy
$$ 
which are derived from the identities for 
homotopy groups 
$\pi_k({\rm Map}_0(S^n,S^m))=\pi_{k+n}(S^m)$
valid for non-negative integers 
$k$ (see, {\it e.g.}, [\tsutsuiA]).  However, the 
two terms are not quite the same 
in status because of the difference in the structure
of connectedness,
$$
\pi_0({\cal C}_1)=\pi_1(S^2)=0\ , 
\qquad
\pi_0({\cal C}_2)=\pi_2(S^2)=\Z\ .
\eqn\zerohomotopy
$$ 
The disconnectedness of the space ${\cal C}_2$
suggests that there are solitons/antisolitons which
may hamper the topological term (\hopfterm) to be
defined firmly.  In fact, this has been pointed out
earlier in [\tsutsuiA, \tsutsuiB] where it was shown that
the Hopf term in the conventional form (\hopfterm)
is well-defined only in the vanishing soliton number sector.

The prime aim of the present paper is to
examine closely the Hopf term in 
the $(2+1)$-dimensional NSM, 
and thereby show that the procedure proposed
in [\tsutsuiA, \tsutsuiB] to extend the applicable 
domain of the term to the whole configuration space
is viable mathematically and also
natural from physical point of view.  
Our argument is based primarily 
on the adjoint orbit parameterization (AOP) of the model
where one uses a group ($SU(2)$-)valued field $g$ 
instead of the vector field $\vecn$ 
in the original description.  
The AOP has been 
introduced in [\BalachandranA, \BalachandranB]
to describe the NSM in the general framework of 
$G/H$ coset models, but it turns out to be 
advantageous over the original description 
especially in treating  
topological quantities, which is a property crucial for 
implementing the procedure proposed.

We also present here a path-integral
derivation of the soliton term (\slterm) --- which is 
well-defined for any configurations 
in the $(1+1)$-dimensional NSM  ---  
by a method similar to that used to induce 
the $\theta$-term 
in Yang-Mills theory in $(3+1)$ dimensions [\Jackiw].
This will highlight the analogy of
the two terms as a quantum mechanically 
induced topological term, 
and at the same time elucidate the usefulness of the 
AOP for the NSM.

We begin our discussion by reviewing 
the Hamiltonian formulation of the model
in $(d+1)$ dimensions using the AOP in section 2.
We then show in section 3 that 
in the $(1+1)$-dimensional NSM
the soliton term arises from the unitary
representations of the fundamental group associated with  
the gauge symmetry inherent to the AOP. 
In section 4, the Hopf term in the $(2+1)$-dimensional NSM
is examined in detail in the original
description, and then in section 5 it is 
re-examined in the AOP to present the procedure of extension
and its physical interpretation.  Section 6 is devoted to
our conclusion and discussions, including 
the possibility of the Hopf term being
a topological `Wess-Zumino term' in $(1+1)$ dimensions.

\secno=2 \meqno=1


\bigskip
\noindent{\bf 2. Hamiltonian formulation in the AOP}
\medskip

The AOP is a description of the NSM identifying
the target space of the field $\vecn(\vecx, t)$ as
a nontrivial adjoint orbit of a Lie group $G$.
For our $O(3)$ NSM, we take $G = SU(2)$ and 
use $\{T_a=\sigma_a/2i; a = 1, 2, 3\}$ 
for a basis of the Lie algebra ${\frak su}(2)$.
With an $SU(2)$-valued field $g(\vecx, t)$ 
we then consider
$$
  g\, T_3\, g^{-1} = n^a T_a,
\eqn\hopffibration
$$
where the l.h.s.~forms the adjoint 
orbit passing through $T_3$,
which is then identified with the target $S^2$ of   
$\vecn(\vecx, t)$ satisfying $|\vecn|=1$.
Note that 
the AOP possesses inherent ambiguity 
associated with the subgroup
$U(1) \subset SU(2)$ generated by the element $T_3$.  Indeed,
$\vecn(\vecx, t)$ is unchanged under the transformations 
$$
  g(\vecx, t) \longrightarrow g(\vecx, t)\, h(\vecx, t), 
\eqn\gaugetrnsf
$$
for any (smooth) function $h(\vecx, t) \in U(1)$.
On account of this ambiguity,  
the boundary condition for $g(\vecx, t)$ which  
corresponds to (\BCone) for $\vecn(\vecx, t)$ becomes
$$
g(\vecx,t) \to g_0 \, k(\vecx,t)
\qquad {\rm as}\quad |\vecx|\to\infty,
\eqn\BCg
$$
where $g_0 \in SU(2)$ is a constant element
fulfilling $ g_0\, T_3\, g_0^{-1} = n_0^a T_a $
and $k(\vecx, t) \in U(1)$ is an arbitrary function 
defined at spatial infinity.
The boundary condition (\BCg) implies that, at a fixed time, 
unlike $\vecn$ which can be 
regarded as a map $S^d \rightarrow S^2$, 
the map $g$ used in the AOP can only 
be treated as a map from a 
$d$-dimensional disc to the group, 
$D^d \rightarrow SU(2)$, with $\partial D^d \simeq S^{d-1}$
being identified with spatial infinity.
Under the AOP the original Lagrangian (\onlsm) becomes\note{%
Conventions: 
The trace is normalized as
$\Tr\ T_aT_b=\delta^{ab}$, and the symbol 
$|_{{\r}}$ denotes 
$X|_{{\r}}=\sum_{a=1}^2 X^aT_a$ 
for $X=\sum_{a=1}^3 X^a T_a$.
Our antisymmetric tensor $\epsilon^{\mu\nu}$ has 
the sign $\epsilon^{01} = 1$, and we use
$\dot u := \partial u/\partial t$ and 
$u' := \partial u/\partial x$ 
(for $d = 1$).}
$$
{\cal L}_0(g)
=\frac{1}{2\lambda^2}\ 
\Tr\ (g^{-1}\partial_{\mu}g|_{{\r}})^2.
\eqn\nlsmtwo
$$
The Lagrangian (\nlsmtwo) has a (trivial) local $U(1)$ 
symmetry under (\gaugetrnsf)
due to the ambiguity of the AOP mentioned above.

In order to construct the Hamiltonian formulation
of the NSM in $(d+1)$-dimensional spacetime, 
we adopt Dirac's procedure for constrained systems
and introduce a set of $SU(2)$ variables
$ \xi^a(\vecx)$, $a = 1, 2, 3$, to parametrize 
$g(\vecx) = g(\xi^a(\vecx))$.  From
their conjugate momenta 
$\pi_a = \partial {\cal L}(g)/\partial \dot{\xi^a}$
and the matrix
$
  N^a_{\ b}T_b := g^{-1}(\partial g/\partial \xi^a),
$
we define
$$
J^a = (N^{-1})^a_{\ b}\ \pi_b=
\cases{
  \frac{1}{\lambda^2}\ N^c_{\ a}\dot{\xi}^c, &$a = 1, 2$, \cr
  0, &$a = 3$,  \cr
}
\eqn\defineJ
$$
with $(N^{-1})^a_{\ b}$ being the inverse of $N^a_{\ b}$,
that is, $(N^{-1})^a_{\ b}\ N^b_{\ c}
=N^a_{\ b}\ (N^{-1})^b_{\ c}=\delta^a_{\ c}$.
The point to be noted is that 
$J(\vecx) = J^a(\vecx) T_a$ 
is independent of the parameterization $\xi^a$
and, together with $g(\vecx)$, 
parametrizes the phase space of the model, forming
the following fundamental Poisson bracket,
$$
\eqalign{
\{\ J^a(\vecx)\ ,\ J^b(\vecy)\ \} &
= -\epsilon^{abc}\,J^c(\vecx)\,\delta^{(d)}(\vecx-\vecy), \cr
\{\ J^a(\vecx)\ ,\ g(\vecy)\ \}
&= g(\vecx)\,T_a\,\delta^{(d)}(\vecx-\vecy), \cr
\{\ g(\vecx)\ ,\ g(\vecy)\ \} &= 0.
}
\eqn\fpb
$$

The Legendre transform of the Lagrangian (\nlsmtwo)
then leads to the Hamiltonian
$H_0(g, J) = \int d^dx\, {\cal H}_0$ with
$$
{\cal H}_0(g, J) = 
\frac{\lambda^2}{2}\ 
\Tr(J|_{{\rm r}})^2+\frac{1}{2\lambda^2}
\ \Tr\ (g^{-1}\partial_ig|_{{\rm r}})^2.
\eqn\hamiltonian
$$
Note that our primary constraint (in Dirac's notation),
$$
\phi(g, J) \approx 0, \qquad \hbox{where} 
\quad \phi(g, J) := J^3(\vecx),
\eqn\generator
$$ 
generates 
the infinitesimal right transformation associated with
the gauge transformation (\gaugetrnsf).
Since the constraint (\generator)
commutes with the Hamiltonian (\hamiltonian) 
under the Poisson bracket (\fpb), no further (secondary) 
constraints arise.

\secno=3 \meqno=1


\bigskip
\noindent{\bf 3. The soliton term}
\medskip

We now specialize to the $(1+1)$-dimensional case
and show that the soliton term (\slterm) arises as
a result of quantization.  To this end, 
we first observe that 
for $d = 1$ the spatial infinity consists of two points
$x = \pm \infty$ and we can always eliminate
the arbitrary function $k(x, t)$ in (\BCg)
by a gauge transformation (\gaugetrnsf).
Thus, instead of (\BCg) we may consider the
simplified boundary condition,
$g(x,t)\to g_0$ as $|x|\to\infty$, without
loss of generality.  Under this condition,
at a fixed time the function
$h(x, t)$ in (\gaugetrnsf) becomes a map from the
(effective) space
$S^1$ to the target $U(1)\simeq S^1$.  
Thus, those gauge transformations can be classified 
by the winding number, 
$$
w(h) = \frac1{4\pi}\int_{S^1} dx\ \Tr\ T_3(h^{-1}h')
\eqn\winding
$$
of the map, and a representative  
map possessing the winding number $n\in\Z$ is given by 
$h_n(x; L) := e^{4n\pi x T_3/L}$,
where $L$ is the length of the space $S^1$.
Gauge transformations with zero winding number 
can be generated by infinitesimal transformations 
and are called `small gauge transformations,' 
whereas those with non-zero winding number are called 
`large gauge transformations.'
Let us next see the consequences of the invariance 
of the theory under these transformations in quantum
theory.

Upon quantizing the model in Schr\"odinger picture, 
state vectors are 
represented by wave functionals $\Phi[g(x)]$
where the argument $g(x)$ denotes a configuration 
at a fixed time.  Observables on phase space
are now regarded as self-adjoint operators 
(although we use the same
notation as before) and, in particular,  
the canonical momenta conjugate to
$\xi^a(x)$ are realized by functional derivatives
$
\pi^a(x)=-i\hbar\delta/\delta\xi^a(x).
$
The gauge symmetry of the theory is ensured
by requiring that under gauge transformations (\gaugetrnsf)
physical functionals be invariant 
$$
\Phi_{{\rm phys}}[g(x)h(x)]=e^{iF[h(x)]}\ 
\Phi_{{\rm phys}}[g(x)],
\eqn\physicalstate
$$
up to a phase $e^{iF}$ given by some functional $F[h(x)]$.
For small gauge transformations this (with $e^{iF} = 1$) 
follows from the condition that implements the first class 
constraint (\generator) in the quantum theory,
$$
J^3(x)\ \Phi_{{\rm phys}}[g(x)]
=(N^{-1})^3_{\ b}\ \frac{\delta}{\delta\xi^b}
\ \Phi_{{\rm phys}}[g(x)]=0,
\eqn\condthree
$$      
which is the analogue of the Gauss' law in gauge theory.  

To find out the phase factor $e^{iF[h]}$ acquired 
for large gauge transformations, let us note that  
any map $h(x)$ having the winding number
$w(h) = n$ can be decomposed
as $h(x) = h_n(x; L) \tilde{h}(x)$ using the 
representative map $h_n(x; L)$ and
$\tilde{h}(x)$ that has zero winding number.  
It then follows from (\condthree) 
that the factor $e^{iF[h]}$ 
depends only on the winding number of $h(x)$, and 
it is given by a unitary representation of
the additive group $\Z$, namely, we have 
$$
\Phi_{{\rm phys}}[g(x)h(x)]
=e^{i\theta w(h)} \Phi_{{\rm phys}}[g(x)],
\eqn\phase
$$
with an arbitrary angle parameter 
$\theta\in[0, 2\pi)$.
The energy eigenstates are obtained by solving
the Schr\"odinger equation,
$H_0(g, J)\,\Phi_{{\rm phys}}=E\,\Phi_{{\rm phys}}$.

For convenience 
we may wish to use states which are invariant
under all gauge transformations.  
This can be accomplished by 
introducing
$$
K(g)=\frac1{4\pi}\int dx\ \Tr\ T_3(g^{-1}g'),
\eqn\no
$$
which transforms under gauge transformations 
(\gaugetrnsf) as
$
K(gh) = K(g) + w(h).
$
Then the desired states 
which are invariant even under 
large gauge transformations can be obtained by
$$
\Psi[g(x)] := e^{i\theta K}
\Phi_{{\rm phys}}[g(x)].
\eqn\no
$$
Noting that 
$
J^ae^{i\theta K}=e^{i\theta K}
(J^a+\frac{\hbar\theta}{4\pi}\ \epsilon^{3ab}\Tr\ T_b(g^{-1}g')),
$
we see that the invariant states $\Psi[g(x)]$
obey the Schr\"odinger equation in the form
$H_\theta(g, J)\,\Psi=E\,\Psi$, where
$H_\theta(g, J) = \int dx\, {\cal H}_\theta$ is the modified
Hamiltonian with
$$ 
{\cal H}_\theta(g, J) = 
\frac{\lambda^2}{2}\left(J^a-\frac{\hbar\theta}{4\pi}
\ \epsilon^{3ab}\Tr\ T_bg^{-1}g'\right)^2+\frac1{2\lambda^2}\ 
\Tr\ (g^{-1}g'|_{{\r}})^2. 
\eqn\no
$$

One can put the above formulation of quantum theory
in the path-integral formalism by means 
of the standard Faddeev-Popov prescription 
(see, {\it e.g.}, [\henneaux]), where one introduces
a gauge fixing condition $\chi(g, J) \approx 0$ 
corresponding to the constraint $\phi(g, J) \approx 0$
in (\generator) and considers the partition function
in phase space,
$$
Z = \int{\cal D}g\, {\cal D}J\,\delta(\phi) \delta(\chi)\,
\vert \det \{\phi, \chi\} \vert\, 
\exp
\left[\ \frac{i}{\hbar}\int d^2x
\left(\Tr\, J(g^{-1}\dot{g})-{\cal H}_\theta\right)\right],
\eqn\partone
$$
with ${\cal D}g\, {\cal D}J 
= \prod_{x, t} [\Tr\ (g^{-1}dg)^3 \prod_a dJ^a]$.  
Choosing the gauge fixing condition to be 
$J$-independent $\chi(g, J) = \chi(g)$, and 
noting that $\{\phi, \chi\}$ gives the 
infinitesimal gauge transformation $\delta \chi$,
one can carry
out the $J$-integrations to get the configuration
space path-integral,
$$
Z = \int{\cal D}g\, \delta(\chi)\,
\vert \det \delta \chi \vert\, 
\exp
\left[\ \frac{i}{\hbar}\int d^2x
\left({\cal L}_0(g) - \frac{\hbar\theta}{8\pi}
\epsilon^{\mu\nu}\partial_\mu
\Tr\, T_3 (g^{-1}\partial_\nu g) \right)\right].
\eqn\pathlag
$$
The first term ${\cal L}_0(g)$ in the 
exponent of the path-integral 
is the Lagrangian (\nlsmtwo), 
whereas the second term is just the soliton term
${\cal L}_{\rm soliton}$ in (\slterm) as can be readily
confirmed upon using (\hopffibration).
It is also obvious 
that the path-integral measure ${\cal D}g\, \delta(\chi)\,
\vert \det \delta \chi \vert$ must be the same as
the measure ${\cal D}\vecn$ for the field $\vecn(x, t)$. 
Indeed, we can see this explicitly 
if we employ the Euler angle decomposition\note{%
Since the decomposition is possible only locally 
in $SU(2)$, for a more rigorous treatment one needs to
introduce a set of patches to cover the $SU(2)$.  
The obstruction
for a na\"{\i}ve global gauge fixing is also evident in 
that in terms of $g$ the soliton term 
is a total divergence while in terms of $\vecn$ it is not.  
}
$g=e^{\alpha T_3}e^{\beta T_2}e^{\gamma T_3}$ 
for which the measure reads 
${\cal D}g=
\prod\sin\beta\ d\alpha\ d\beta\ d\gamma={\cal D}\gamma
{\cal D}\vecn$, 
and whereby choose $\chi(g) = \gamma$ which has
$\delta \chi = \hbox{const}$.  
We therefore arrive at
$$
Z=\int{\cal D}\vecn\ 
\exp\left[\ \frac{i}{\hbar}\int d^2x
\left(\frac1{2\lambda^2}\ (\partial_\mu\vecn)^2+
\frac{\hbar\theta}{8\pi}\ \epsilon^{\mu\nu}\epsilon_{abc}
n^a\partial_\mu n^b\partial_\nu n^c\right)\ \right],
\eqn\no
$$
which shows that the soliton term (\slterm) is induced
upon quantization in the $O(3)$ NSM in $(1+1)$ dimensions.

\secno=4 \meqno=1


\bigskip
\noindent{\bf 4. The Hopf term in ${\vecn}(\vecx, t)$}
\medskip

The Hopf term (\hopfterm) has been widely used 
in the physics literature especially in the context of
fractional spin and statistics in $(2+1)$ dimensions.
However, unlike the soliton
term in $(1+1)$ dimensions,
the Hopf term is 
not well-defined mathematically for 
generic configurations, and hence it requires a careful 
consideration before it is used.
To examine the problem in detail, we first recall 
the Hopf invariant used in mathematics
(see, {\it e.g.}, [\Bott]).
Let $f$ be a map $S^3 \rightarrow S^2$ and
$F$ be a generator of the de Rham cohomology 
$H^2_{\rm DR}(S^2) = \R$.  Since $H^2_{\rm DR}(S^3) = 0$
the pullback $f^*F$ of $F$ under the map $f$
({\it i.e.}, $F$ regarded as a 2-form on $S^3$) 
admits the form
$f^*F = dA$ with some 1-form
$A$ on $S^3$.  The Hopf invariant associated to the map $f$ 
is then given by 
$$
H(f) := - \frac{1}{16\pi^2} \int_{S^3}A\wedge dA\ ,
\eqn\HopfTerm
$$
where the normalization is chosen such that $H(f) \in \Z$
when $F$ is normalized as $\int_{S^2} F \in 4\pi \Z$.
Note that the topological invariant $H(f)$ in (\HopfTerm) 
is independent of the choice of $A$.  
In other words, despite that 
there exists an ambiguity in $A$ (under $U(1)$
gauge transformations $A \rightarrow A - d\Lambda$) and 
hence in the integrand in (\HopfTerm), the integral
is still uniquely determined.
With an appropriately normalized
$F$ the Hopf invariant $H(f)$ becomes an 
integer characterizing the map $f$. 

Evidently, if our vector field $\vecn$ can be regarded
as a map $S^3 \rightarrow S^2$ with $S^3$ our (effective) 
spacetime, then by identifying $\vecn$ with
the above $f$ and also 
${1\over 2}F_{\mu\nu} dx^\mu\wedge dx^\nu$ 
with the 2-form $f^*F$, the integral of the Hopf term
(\hopfterm) over the spacetime $S^3$
becomes identical to (\HopfTerm) up to an overall constant 
and, therefore, it is well-defined.  
However, the boundary
condition in space (\BCone) implies that we shall be 
considering the NSM in a spacetime of the form
$M = S^2 \times I$ where $I$ is a time interval, say,
$I = [0, T]$, or $M = S^2 \times S^1$ if
an additional periodic boundary condition in time is imposed.
Thus in general our map is given by 
$\vecn: M \rightarrow S^2$ with these $M$, for which we have
$H^2_{\rm DR}(M) \ne 0$.  Accordingly, there is no
guarantee to find such (globally defined) 
$A$ satisfying $\vecn^*F = dA$ to a given map $\vecn$.  
In other words,         
the na\"{\i}ve integral,
$$
I(A; M) := - \frac{1}{16\pi^2} \int_M A\wedge dA,
\eqn\csterm
$$
cannot be used for providing the topological term we want.
At this point one may think that a solution is to regard 
$A$ as a $U(1)$ connection over
the base space $M$, that is, to find $A$ given locally 
on patches which are introduced to trivialize the fibre bundle.
That this does not work can be seen as follows.

Suppose $M$ is covered by two local patches, 
$M_1$ and $M_2$, on which we have 
a 1-form $A_1$ and $A_2$, respectively, 
satisfying $\vecn^*F = dA_a$ for $a = 1$, 2.
One then may consider, instead of (\csterm), 
the sum of integrals\note{%
One can also use $A_2$ instead of $A_1$ in the
last term, and this creates another ambiguity 
in defining the topological term.}
$$
I(A_1; M_1) + I(A_2; M_2) - I(A_1; M_1\cap M_2).
\eqn\MHopfTerm
$$
This is, however, not 
invariant under gauge transformations 
$A_a \rightarrow A_a - d\Lambda_a$ performed 
separately on the patches with $\Lambda_1$ and 
$\Lambda_2$ chosen independently, because
its variation, $\int_{\partial M_1} \Lambda_1 dA_1$ plus 
contributions from other two terms, does not vanish
for generic $\Lambda_a$.
(The gauge noninvariance of the conventional Hopf term  
has been pointed out earlier in [\OS].)
Actually, this is a problem with the Chern-Simons term
in gauge theory (rather than with the Hopf term in the NSM),
where it is assumed that the connection $A$ is globally
defined [\Deser], which is enough if one is interested in
the perturbative analysis of the theory.  
The real problem with the Hopf term is that
in the $O(3)$ NSM this assumption excludes solitons 
or anti-solitons unless the total charge vanishes, 
which are considered to be responsible for
physical phenomena of our interest such as fractional spin.
Indeed, for the spacetime $M = S^2 \times I$
(or $M = S^2 \times S^1$) the assumption   
amounts to the requirement that the 2-form $\vecn^*F$ 
be a trivial element of $H^2_{\rm DR}(M)$ and, therefore, 
we find that the soliton charge,
$$
Q(\vecn) := -\frac{1}{4\pi}\int_{S^2} \vecn^*F \ ,
\eqn\solitoncharge
$$
becomes
$Q(\vecn) = -(1/4\pi)\int_{S^2} d A = 0$, 
that is, the map $\vecn$ must belong to the sector 
where the soliton charge vanishes.

Conversely, it is possible to show
that the restriction to the sector $Q(\vecn) = 0$ is
sufficient 
for the topological term to be well-defined in the NSM.  
For this, we first note that, for a generic
spacetime $M$, 
the condition $H^2_{\rm DR}(M)=0$
is not still enough for ensuring the integral (\csterm)
to be well-defined, because 
the gauge invariance of the integral (\csterm) requires
that the 1-form $A$ must in general be gauge-fixed on 
$\partial M$ if $\partial M \ne 0$.
In quantum theory, however, such a 
gauge fixing is not necessary
with our space time $M = S^2 \times I$.  
The reason for this is that, as is well-known,
in quantum theory 
we can always place periodic boundary conditions
in time,\note{%
This is easily 
realized in the path-integral for a particle, 
where one needs to define
a {\it relative} phase to a given pair of 
paths, rather than to define
an {\it absolute} phase to each path, in order to provide
the transition amplitude.  The relative
phase between a pair of two paths obeying the
same boundary conditions at $t=0$ and $T$ 
can be regarded as the phase attached to the path given
by connecting the two paths (with the time of 
one of the paths reversed).  This amounts to
defining a phase to an arbitrary loop which is a path
possessing the same initial and final points.
}
and this allows us to put the time
period $I$ into $S^1$ and thereby remove the boundary of 
the spacetime.    
We then notice that if
the aforementioned restriction to the $Q(\vecn) = 0$ sector
is made in the NSM, we can deform any map
$\vecn(\vecx)$ at $t = 0$ and $T$ continuously
to the constant map $\vecn(\vecx) = \vecn_0$ without changing
the integral in (\csterm) 
(since we have $\partial(S^2 \times S^1) = 0$).
This procedure allows us to regard $\vecn$ as a map 
$S^3 \rightarrow S^2$ by shrinking the space $S^2$ to a point 
at the both ends of time and, consequently,  
the integral in (\csterm), which is just
the Hopf invariant (\HopfTerm), can be used as a 
topological term in the model.\note{%
In the literature [\Bowick, \Semenoff, \Forte, \DHokerB]
the term (\csterm) is considered
even for sectors with nonzero soliton numbers
under the Coulomb gauge.
However, as has been pointed out in [\tsutsuiB],
the fractional spin evaluated directly from the term 
becomes gauge dependent and physically unacceptable.
}
%

\secno=5 \meqno=1


\bigskip
\noindent{\bf 5. The Hopf term in $g(\vecx, t)$}
\medskip

The foregoing argument shows that there are basically
three obstacles for the na\"{\i}ve integral
$I(A; M)$ in (\csterm)
to be well-defined as 
the topological term associated with the Hopf invariant
(\HopfTerm).  These obstacles are related to the 
conditions that the term be
(i) well-defined as an integral over the spacetime $M$, 
(ii) a topological invariant, and 
(iii) gauge invariant.
As we shall see shortly, in the AOP 
the situation concerning these conditions 
turns out to be quite different, and we shall  
exploit it in order to define the topological term
to any configurations, fulfilling all the three conditions.

To this end, we first note that 
the corresponding $F$ becomes then the 2-form 
on the target of the map 
$g: D^2 \times I \rightarrow SU(2) \simeq S^3$,
and it is given by $F = -\Tr T_3(g^{-1}dg)^2$.
Since $H_{\rm DR}^2(D^2 \times I) = 0$ 
we always have a 1-form $A$
such that the pullback of $F$ under $g$ becomes $g^*F = dA$ 
--- an obvious solution is 
$A = \Tr T_3(g^{-1}dg)$.  Thus we now have 
$A$ local in the variable $g$ in contrast
to the previous $A$ which is nonlocal in $\vecn$.
Accordingly, in the AOP the integral (\csterm) turns into
the local expression [\BalachandranA],
$$
I(A; M) =
I(g) := \frac{1}{48\pi^2}\int_{\bar M} \Tr(g^{-1}dg)^3\ ,
\eqn\wdn
$$
where now $\bar M = D^2 \times I$, and this
shows that condition (i) is fulfilled. 
In passing 
we mention that, on account of the boundary condition
(\BCg), the AOP version of the soliton charge 
(\solitoncharge) reads 
$$
Q(g) := - \frac{1}{4\pi}\int_{D^2} g^*F 
= - \frac{1}{4\pi}\int_{\partial D^2} \Tr T_3(g^{-1}dg)\ , 
\eqn\no
$$ 
which gives the winding number of the map
$k: {\partial D^2} \simeq S^1 \rightarrow U(1)$ 
at space infinity.

Let us now observe that, when the spacetime $\bar M$ 
for $g$ can be regarded as $D^2 \times S^1$
as can be done in the path-integral,  
the integral (\wdn) becomes a topological
invariant irrespective of the soliton sector we are in.
Indeed, under an arbitrary variation 
$g \rightarrow g + \delta g$ we obtain
$$
I(g + \delta g) - I(g) = \frac{1}{16\pi^2}
\int_{\partial \bar M} 
\Tr g^{-1}\delta g(g^{-1}d g)^2\ ,
\eqn\variation
$$
which vanishes for any $g$ which are 
assumed to be of the form (\BCg) at 
$\partial \bar M = \partial D^2 \times S^1$ and hence
$(g^{-1}d g)^2\vert_{\partial \bar M} 
= (k^{-1}d k)^2\vert_{\partial \bar M} = 0$.
However, the problem is that $I(g)$ is not
invariant under time-dependent gauge transformations
(\gaugetrnsf)
possessing nontrivial winding numbers along $S^1$.
This can be confirmed directly by observing
$$
I(gh) - I(g) = 
- \frac{1}{16\pi^2}\int_{\partial \bar M} 
\Tr(k^{-1}d k)\wedge (dh\, h^{-1})\ .
\eqn\vargauge
$$
To evaluate the r.h.s.~of (\vargauge), let $L$ be
the length of the boundary $\partial D^2 \simeq S^1$ 
and regard the domain of the integral 
$\partial \bar M = S^1 \times S^1$
as the rectangle $I^2 = [0, L] \times [0, T]$ in which
periodic boundary conditions are imposed on the maps 
$k$ and $h$.  Introducing the
coordinates $(x, t) \in I^2$, we put
$$
k(x, t) = e^{\xi(x, t)T_3}\,h_m(x; L)\, h_l(t; T)\ ,
\qquad
h(x, t) = e^{\eta(x, t)T_3}\, h_n(t; T)\ ,
\eqn\explicit
$$
where $\xi(x, t)$ and $\eta(x, t)$ are periodic functions
in $I^2$, $h_m(x; L)$, $h_l(t; T)$, $h_n(t; T)$ are 
the representative maps defined earlier with $m$, $n$, $l \in \Z$.
Note that the integer $m$ in $k(x, t)$ 
equals (minus) the soliton number $- Q(g)$, 
whereas for $h(x, t)$ no such integer appears
since we have $Q(h) = 0$ on account
of the fact that for gauge transformations $h$ must be given
on $D^2$ which is contractible.
Substituting (\explicit) into (\vargauge) we find
$$
I(gh) - I(g) = 
- \left(\frac{1}{16\pi^2}\right)\left(\frac{4n\pi}{T}\right)
  \left(\frac{4m\pi}{L}\right)
\int_{I^2} dx \wedge dt 
= n\, Q(g)\ ,
\eqn\difhopf
$$
that is, condition (iii) is not fulfilled unless $Q(g) = 0$.

In order to extend the domain of $I(g)$ to 
nonvanishing soliton number sectors,  
a possible procedure proposed in [\tsutsuiB] 
is that one imposes a (partial) gauge
fixing condition on $g(\vecx, t)$ such that at spatial infinity
it become time-independent, {\it i.e.}, 
$$
{{\partial}\over{\partial t}} g(\vecx, t) 
\Bigr\vert_{\partial D^2} = 0\ .
\eqn\weylgauge
$$ 
Indeed, this excludes those gauge transformations
by $h(x, t)$ having $n \ne 0$ on $\partial D^2$, 
and hence
the r.h.s.~of (\difhopf) vanishes irrespective of the soliton
charge $Q(g)$.  Thus we see that, upon imposing (\weylgauge),
the integral $I(g)$ defines a topological invariant and,
at the same time, it is gauge invariant for any $g$.  
It remains, therefore, 
to find the meaning of the topological invariant so defined.

For this, let us consider the configuration,
$$
\bar g(\vecx, t) := g_0\, 
g^{-1}(\vecx, 0) \, g(\vecx, t)\ .
\eqn\conv
$$
Using the additivity of the soliton number,
$Q(g_1 g_2) = Q(g_1) + Q(g_1)$ for any $g_1$, $g_2$
obeying (\BCg), we find that $Q(\bar g) = 0$.  
Thus we are allowed to regard $\bar g$ as 
one converted from $g$ to the vanishing soliton number 
sector, for which the integral $I(\bar g)$ in 
(\wdn) gives the Hopf invariant without imposing 
(\weylgauge).
On the other hand, it can be readily shown [\tsutsuiB] that, 
once the condition (\weylgauge) is imposed, the conversion
does not change the integral, $I(g) = I(\bar g)$.  
It thus follows that
the topological invariant defined above for any $g$ with 
(\weylgauge) is nothing but the Hopf invariant 
for the converted configuration $\bar g$.  
\topinsert
\vskip 1cm
\let\picnaturalsize=N
\def\picsize{5.5cm}
\def\picfilename{fig.epsf}
\ifx\nopictures Y\else{\ifx\epsfloaded Y\else\input epsf \fi
\global\let\epsfloaded=Y
\ifx\picnaturalsize N\epsfxsize \picsize\fi
\hskip 2.5cm\epsfbox{\picfilename}}\fi
\vskip 0cm
\abstract{%
{\bf Figure 1.} 
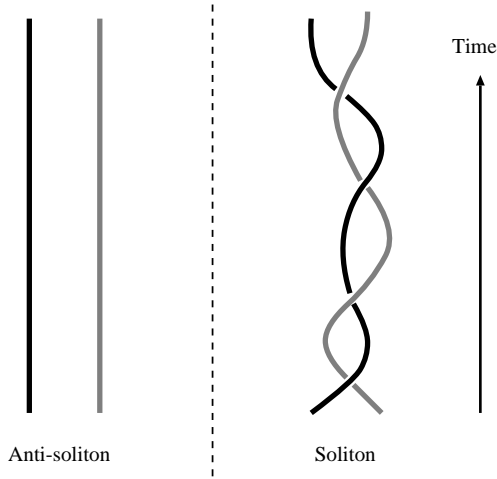
The physical picture of the conversion procedure.  
Given a soliton configuration, which evolves in time and is
shown on the right, the procedure puts a static anti-soliton
on the left in a distance far from the soliton, so that
the entire soliton charge vanishes.  The Hopf number will be
gained by the dynamical evolution of the soliton.
}
\endinsert

Note that conversion to the vanishing soliton number sector
is far from unique.  However, since the integral (\wdn)
is a topological invariant, any conversion of the form,
$$
\bar g(\vecx, t) :=  
g_{\rm A}(\vecx) \, g(\vecx, t)\ ,
\eqn\newconv
$$
gives the same value for $I(\bar g)$ as long as
the static configuration $g_{\rm A}$ has the soliton
number opposite to that of $g$ so that $Q(\bar g) = 0$
(and we may also 
require $g_{\rm A}\vert_{\partial D^2} = 1$
so that $\bar g$ still obeys (\BCg)).
This observation allows for the following physical
interpretation to the topological invariant we just 
assigned to $g$.  Suppose that the original 
configuration $g(\vecx, t)$ has soliton number $n$,
and that it is localized in some finite domain in space.
Choose then $g_{\rm A}(\vecx)$ such that it has 
soliton number $(-n)$ and also localized in some
other domain which do not intersect with the domain of $g$.
Then what the conversion (\newconv) is doing is that
it places a static anti-soliton $g_{\rm A}$ 
somewhere far from the soliton $g$.  Roughly speaking, 
the Hopf invariant counts the number of twists made by
the configuration during the time interval $[0, T]$,
and for $\bar g$ these twists are performed by
the soliton part.  The static anti-soliton does not
play a role in this, except that it provides a ground
in which the Hopf invariant becomes well-defined by
nullifying the soliton charge (see Fig.~1).
In fact, 
the physical picture originally 
used by Wilczek and Zee [\Wilczek] to discuss
fractional spin (for the case $n = 1$) is 
the initial soliton/anti-soliton pair creation  
and annihilation at the final time,
which is the case where the conversion is made
according to (\conv) for a soliton $g(\vecx, t)$ 
which stays far from the static anti-soliton 
$g_{\rm A}(\vecx) = g^{-1}(\vecx, 0)$ except when the
creation and annihilation take place at $t = 0$ and $t = T$,
respectively. 
Our procedure presented here provides a 
mathematical ground for the picture,
in the light of finding a well-defined topological invariant
corresponding to the Hopf invariant.  We also point out that
it is possible to carry out the procedure in the original 
description in terms of $\vecn$, but it becomes more involved
than the one given here which exploits fully the advantage
of the group properties of the AOP.

\secno=6 \meqno=1


\bigskip
\noindent{\bf 6. Conclusion and discussions}
\medskip

In this paper we discussed two 
types of topological terms in the
$O(3)$ NSM, one is the soliton
term in $(1+1)$ dimensions and the other 
is the Hopf term in $(1+1)$ dimensions.  
In contrast to the soliton term which is well-defined
and can be derived from the unitary representation of the
fundamental group of the configuration space ${\cal C}_1$, 
the Hopf term used in the literature is ill-defined
and, against the general expectation, 
it cannot serve to produce
fractional spin and statistics in its conventional form.
We argued that the conversion procedure, which has been 
proposed earlier to make the Hopf term well-defined
and is equivalent to a partial gauge fixing in the AOP,
is natural both mathematically and physically.

Once we constructed the Hopf term as a topological 
invariant associated with the
first homotopy group (\firsthomotopy), 
we may ask if it can also be used as 
a `Wess-Zumino term' in the
$(1+1)$-dimensional NSM.  This  
possibility arises from the fact
that the second homotopy group of 
the configuration space ${\cal C}_1$ reads
$$
\pi_2({\cal C}_1)=\pi_3(S^2)=\Z,
\eqn\secondhomotopy
$$
which suggests that the NSM may admit an 
associated topological term 
analogous to the Wess-Zumino term in 
the Wess-Zumino-Novikov-Witten model [\Witten]
which has the same second homotopy group.
In fact, the usual Wess-Zumino term is 
given precisely by the integral (\wdn)
with $M = D^3$ whose boundary $\partial M \simeq S^2$
is identified with the $(1+1)$-dimensional spacetime. 
Since in the NSM the second homotopy group is related to
the Hopf fibration, we may expect that 
the topological term is again the Hopf term 
we have just made well-defined.

To examine this possibility, 
for definiteness we take our spacetime to be $S^2$, 
which is always possible since by (\zerohomotopy) 
there is no obstacle 
to deform any configurations to a constant one.
We however observe that taking the extrapolated manifold
$M = D^3$ for our Wess-Zumino term (\csterm)
is not possible, because
a loop in ${\cal C}_1$ given by the spacetime map $\vecn(x, t)$
cannot always be deformed to a point due to (\firsthomotopy),
unlike in the Wess-Zumino-Novikov-Witten model 
where it can.  Instead,
we may consider $M = S^2\times I$
parameterized by $(x, t, \sigma)$ with the extrapolation
parameter $\sigma\in[0,1]$ in such a way that 
the map $\vecn(x, t, 1) := \vecn(x, t)$ be extended to
$\vecn(x, t, 0)$ given by 
some fixed configuration possessing the same soliton number
as $\vecn(x, t)$.  This does not render the
integral (\csterm) well-defined either, since the integral 
changes under
gauge transformations for $A$ yielding a
gauge dependent integral over the spacetime $S^2$.
The situation is unaltered even if we take any
three-dimensional manifold for $M$ as long as it 
contains $S^2$ in the boundary $\partial M$, which
is a quality requisite to a Wess-Zumino term 
in $(1+1)$ dimensions.  We also note that, despite
the similarity of the present problem with that of
defining the Hopf term in $(2+1)$ dimensions, 
the procedure we adopted in the AOP cannot be employed
here, because 
one cannot use the periodic boundary condition
in the direction of the extrapolating 
parameter $\sigma$,
as done in the previous case where the role of 
$\sigma$ is played by the time $t$.

Finally, we mention that the construction of
gauge invariant topological terms as a
Wess-Zumino term has also been 
discussed elsewhere [\dhoker] in the 
context of coset models, where 
the list of cohomology generators 
of a symmetric space $G/H$ is exhausted.  However, 
this amounts to finding a local gauge invariant
integrand such as $F\wedge F$ or $F\wedge *F$,
rather than directly seeking for a gauge invariant integral 
as we did above
without assuming the locality and gauge invariance of
its integrand.  The conclusion of the non-existence of such a term, 
however, remains the same.


\ve
\baselineskip= 15.5pt plus 1pt minus 1pt
\parskip=5pt plus 1pt minus 1pt
\tolerance 8000
\vfill\eject

  \vfill\eject\immediate\closeout\reffile
  \centerline{{\bf References}}\bigskip\frenchspacing%
  \input refs.tmp\vfill\eject\nonfrenchspacing
\bye